\documentclass[10pt,twocolumn]{IEEEtran}
\usepackage{graphicx}
\usepackage{epsfig,amsfonts,amsmath,amssymb,amstext,amsthm}
\usepackage{threeparttable}
\usepackage{epstopdf}
\usepackage{footnote}
\usepackage{multirow}
\usepackage{setspace}

\newcommand{\bm}[1]{\mbox{\pmb{$#1$}}}

\newcommand{\dref}[1]{(\ref{#1})}
\newcommand{\Tr}{\mbox{Tr}}

\begin{document}

\title{The Effect of Macrodiversity on the Performance of Maximal Ratio Combining in Flat Rayleigh Fading}

\author{Dushyantha A. Basnayaka,~\IEEEmembership{Student Member,~IEEE,}
        Peter J. Smith,~\IEEEmembership{Senior Member,~IEEE}
        and~Philippa A. Martin,~\IEEEmembership{Senior Member,~IEEE}% <-this % stops a space
\thanks{D. A. Basnayaka, P. J. Smith and P. A. Martin are with the Department of Electrical and Computer Engineering, University of Canterbury, Christchurch, New
Zealand. E-mail:\{dush, p.smith, p.martin\}@elec.canterbury.ac.nz.}
\thanks{D. A. Basnayaka is supported by a University of Canterbury International Doctoral Scholarship.}}% <-this % stops a space

\maketitle
\begin{abstract}
The performance of maximal ratio combining (MRC) in Rayleigh
channels with co-channel interference (CCI) is well-known for
receive arrays which are co-located. Recent work in network MIMO,
edge-excited cells and base station collaboration is increasing
interest in macrodiversity systems. Hence, in this paper we consider
the effect of macrodiversity on MRC performance in Rayleigh fading
channels with CCI. We consider the uncoded symbol error rate (SER)
as our performance measure of interest and investigate how different
macrodiversity power profiles affect SER performance. This is the
first analytical work in this area. We derive approximate and exact
symbol error rate results for $M$-QAM/BPSK modulations and use the
analysis to provide a simple power metric. Numerical results,
verified by simulations, are used in conjunction with the analysis
to gain insight into the effects of the link powers on performance.
\end{abstract}


\section{Introduction}\label{sec:introduction}
Maximum ratio combining (MRC) is a well-known linear combining
technique that maximizes the signal-to-noise ratio (SNR) in noise
limited systems \cite{SiAl00}. In the presence of co-channel
interferers, MRC is sub-optimal compared to minimum mean squared
error (MMSE) combining. However, MMSE combining requires
instantaneous channel knowledge of both the desired source and
interfering sources. In contrast, MRC only requires a knowledge of
the desired source and hence is simpler to implement. For this
reason, there is still interest in MRC processing in the presence of
interference. In \cite{Ngo11}, MRC is investigated for large systems
where it was shown that in the limit as the number of antennas
increases, intercell interference effects disappear. In
\cite{Zhu08}, a switched MRC/MMSE receiver is proposed where the
simplicity of MRC is preferred when the interference levels drop
below a threshold. Here, MRC performance in the presence of small
but non-zero interference is important. There are well-known methods
to estimate the interference level in comparison with the signal
level as described in \cite{PaBe00}.\\
The performance of MRC systems with co-located antenna arrays is
well known for Rayleigh fading channels with multiple co-channel
interferers \cite{Cui99, Tok06}. Recently, interest in distributed
combining has grown due to research in cooperative systems,
base-station collaboration \cite[pp.~69]{Big07}, edge-excited cells
\cite{Catreux99,Zhang10} and network MIMO \cite{Siva07, Fosch06}. In
the standards, distributed processing is part of coordinated
multipoint transmission (CoMP) in LTE Advanced. For these
macrodiversity systems, every link may have a different average SNR
since the sources and the receive antennas are all in different
locations. This variation in SNR makes performance analysis more
complex and to the best of our knowledge no analytical results are
currently available for such systems.\\
Hence, in this paper we analyze the symbol-error-rate (SER) of
macrodiversity MRC systems. Note that the system is not new.
Standard MRC processing is considered and so the general form of the
receiver output and the initial steps in the performance evaluation
are well-known. However the macrodiversity layout creates a new
channel structure which is far more complex than the microdiversity
channel. Hence, the MRC output has a completely new statistical
distribution and a novel, more advanced analysis is required for
system performance evaluation. In particular, we consider a
distributed antenna array performing MRC combining for a single
antenna desired source in the presence of an arbitrary number of
single antenna co-channel interferers. The analysis also covers the
case where both the desired and interfering sources may have
multiple antennas. Since the sources and the receive antennas are
not co-located, the channels are normally independent and so the
focus is on independent Rayleigh fading channels where each link has
a different SNR. In this paper, we evaluate the SER over Rayleigh
fading for fixed values of the long term link SNRs. Hence, the SER
is computed over fast fading while path loss effects and shadowing
are held constant. Looking at the joint effects of the slow fading
(see, for example, \cite{MaChKa10, ZhWoJi09}) would be an
interesting topic for future work. In the scenario where some
sources have multiple antennas, there may be spatial correlation in
the channels corresponding to the antennas at that source. However,
this is beyond the scope of the current work where independent
channels are considered. We provide specific results for BPSK and
QPSK modulations, but the analysis can be applied to $M$-QAM and a
wide range of modulations where the SER can be written in terms of
an expected value of the Gaussian $Q$-function and $Q^2$-function.
The general analytical approach follows the techniques in
\cite{Proakis01}. The novelty in the analysis is the identification
of a representation for the interference and noise term in the
combiner output and the use of this representation in exact SER
calculations. We then use the SER results to analyze the effect of
macrodiversity on MRC performance.

The rest of the paper is organized as follows. In Sec. II, we give
the system model and in Sec. III the performance results and SER are
derived. Sec. IV gives numerical results where the analysis is
verified by simulation and conclusions are presented in Sec. V.

\section{System Model}\label{sec:system_model}
\begin{figure}[!t]
\centering
\includegraphics[scale=0.60]{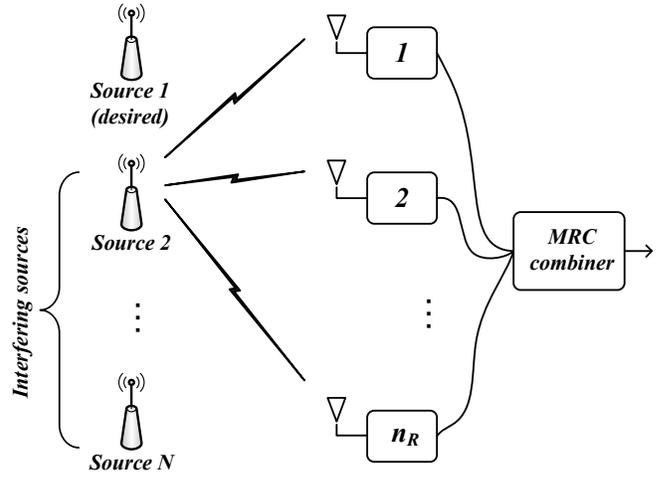}
\noindent \caption{System diagram. To reduce clutter, only paths
from source 2 are shown.} \label{fig:mrc:fig1}
\end{figure}

Consider $N$ single-antenna distributed users communicating with
$n_R$ distributed transmission points (TP) \cite{DaBrEr12} each with
a single receive antenna over an independent flat fading Rayleigh
channel. The system diagram is given in Fig.\ref{fig:mrc:fig1}. The
received signal is given by
\vspace{0mm}
\begin{eqnarray}
\bm{r} &=& \bm{Hs} + \bm{n}, \label{eq:mrc:main1}
\end{eqnarray}
\noindent where $\bm{r}=\left(r_1, r_2, \dots, r_{n_R} \right)^T$ is
the $\mathcal{C}^{n_R \times 1}$ receive vector,
$\bm{H}=\left(h_{ik}\right)$ is the $\mathcal{C}^{n_R \times N}$
channel matrix, $\bm{s}=\left ( s_1 , s_2, \dots, s_{N} \right )^T$
is the $\mathcal{C}^{N \times 1}$ signal vector and
$\bm{n}=\left(n_1, n_2, \dots, n_{n_R} \right)^T$ is the
$\mathcal{C}^{n_R \times 1}$ additive-white-Gaussian-noise (AWGN)
vector at the receive antennas such that $\bm{n}\sim
\mathcal{CN}\left ( \bm{0}, \sigma^2 \bm{I} \right )$. The signals
are normalized to be zero-mean, unit power variables so that $E
\left \{ |s_i|^2 \right \}=1$ for $i=1,2, \dots, N$. The channel
matrix, $\bm{H}$, has independent zero-mean, complex Gaussian
elements such that $E \left \{ |\bm{H}_{ik}|^2 \right \}= P_{ik}$.
Hence, equation \dref{eq:mrc:main1} can be rewritten as
\vspace{0mm}
\begin{eqnarray}
\bm{r} &=& \left( \bm{H}_w \circ \bm{P}^{\circ \frac{1}{2}} \right)
\bm{s} + \bm{n}, \label{eq:mrc:main11}
\end{eqnarray}
\noindent where $\bm{P}=\left(P_{ik}\right)$, $\bm{P}^{\circ
\frac{1}{2}}$ is the element-wise square root of $\bm{P}$, the
operator, $\circ$, represents Hadamard multiplication and the
elements, $\bm{H}_{w,ik}$, of $\bm{H}_w$ satisfy $\bm{H}_{w,ik} \sim
\mathcal{CN}\left ( 0,1\right ) \quad \forall i, k$. The matrix,
$\bm{P}$, is the global power matrix for the system and for the
$k^{th}$ source, an individual power matrix is also defined by
$\bm{P}_k= E \left\{\bm{h}_k \bm{h}_k^H\right\}=\mbox{diag}\left(
P_{1k}, P_{2k}, \dots, P_{n_Rk} \right)$, for $k=1,2, \dots, N$. In
the microdiversity case, $\bm{P}_k \propto \bm{I}$. In
macrodiversity scenarios, $\bm{P}_k$ is no longer proportional to
the identity and these more general power matrices make the analysis
more complex. Assume, without loss of generality, that user 1 is the
desired user. For the purpose of decoding user 1,
\dref{eq:mrc:main1} can be rewritten as
\begin{align}
\bm{r} &= \bm{h}_1 s_1 + \tilde{\bm{H}}\tilde{\bm{s}} + \bm{n} \\
&= \bm{h}_1 s_1 + \bm{i}, \label{eq:mrc:main2}
\end{align}
\noindent where $\bm{h}_1$ is the first column of $\bm{H}$,
$\tilde{\bm{H}}$ is all columns of $\bm{H}$, excluding the first
column, meaning $\bm{H}=\left(\bm{h}_1, \tilde{\bm{H}}\right)$, and
$\tilde{\bm{s}}=\left (s_2, \dots, s_{N} \right)^T$. The $n_R \times
1$ vector, $\bm{i}$ is the interference and noise vector. With MRC
processing, the output of the combiner is given by \cite{Proakis01}
\begin{equation}
\tilde{\bm{r}} = \frac{\bm{h}_1^H \bm{r}}{\bm{h}_1^H \bm{h}_1} = s_1
+ \frac{\bm{h}_1^H \bm{i}}{\bm{h}_1^H \bm{h}_1}.
\label{eq:mrc:main3}
\end{equation}

\noindent The interference and noise term in \dref{eq:mrc:main3} can
be written as
\begin{equation}
Z = \frac{\bm{h}_1^H \bm{i}}{\bm{h}_1^H \bm{h}_1} = \frac{\bm{h}_1^H
\left( \tilde{\bm{H}} \tilde{\bm{s}} + \bm{n} \right)}{\bm{h}_1^H
\bm{h}_1}. \label{eq:mrc:main4}
\end{equation}
Following the standard approach \cite{Proakis01}, we develop a
conditional Gaussian representation for $Z$ as follows. Since
$\tilde{\bm{H}}$ and $\bm{n}$ are zero-mean Gaussian and independent
of $\bm{h}_1$ and $\tilde{\bm{s}}$, it follows that $Z$ is also
zero-mean Gaussian conditioned on $\bm{h}_1$ and $\tilde{\bm{s}}$.
The conditional variance of $Z$ is given by
\begin{align}
\small{ E \! \left \{ \! \left| Z \right|^2 |
\bm{h}_1,\tilde{\bm{s}} \right \}} \! &=\! \small{ E \! \left \{\!
\! \left. \frac{\bm{h}_1^H \left (\tilde{\bm{H}}\tilde{\bm{s}} \!+\!
\bm{n} \right)\left (\tilde{\bm{s}}^H \tilde{\bm{H}}^H \!+\!
\bm{n}^H
\right) \bm{h}_1}{ \left(\bm{h}_1^H \bm{h}_1 \right)^2 } \right| \bm{h}_1, \tilde{\bm{s}} \right \},} \\
&= \frac{\bm{h}_1^H E \left \{ \tilde{\bm{H}}\tilde{\bm{s}}
\tilde{\bm{s}}^H\tilde{\bm{H}}^H + \sigma^2
\bm{I} \right \} \bm{h}_1}{\left(\bm{h}_1^H \bm{h}_1 \right)^2 }, \\
&= \frac{\bm{h}_1^H \left ( \sum_{k=2}^N \bm{P}_k \left|s_k
\right|^2 + \sigma^2 \bm{I} \right) \bm{h}_1}{\left(\bm{h}_1^H
\bm{h}_1 \right)^2 }. \label{eq:mrc:z}
\end{align}
\noindent Hence,  since $Z$ is a conditional Gaussian with variance
given by \dref{eq:mrc:z}, it follows that $Z$ has the exact
representation
\begin{align}
Z &= \frac{\sqrt{\bm{h}_1^H \left ( \sum_{k=2}^N \bm{P}_k
\left|s_k\right|^2 + \sigma^2 \bm{I} \right)
\bm{h}_1}}{\left.\bm{h}_1^H \bm{h}_1 \right.} U,
\label{eq:mrc:main5}
\end{align}
\noindent where $U \sim \mathcal{CN}\left(0,1\right)$. Using this
representation in \dref{eq:mrc:main3} gives the combiner output in
simplified signal plus noise form as
\begin{align}
\tilde{\bm{r}}= s_1 + \frac{\sqrt{Y}}{X}U, \label{eq:mrc:main6}
\end{align}
\noindent where $X=\bm{h}_1^H \bm{h}_1$, $Y=\bm{h}_1^H
\bm{D}\left(\tilde{\bm{s}}\right)\bm{h}_1$ and
$\bm{D}\left(\tilde{\bm{s}}\right)=\sum_{k=2}^N \bm{P}_k
\left|s_k\right|^2 + \sigma^2 \bm{I}$.
\section{Performance Analysis}\label{sec:perform_analysis}

\subsection{A Simple SER Analysis}\label{sub_sec:simple_SER_analysis}
With the combiner output given by \dref{eq:mrc:main6}, SERs for many
modulations can be obtained using standard methodology
\cite{Proakis01}. As an example, for BPSK, we have the SER
\begin{align}
P_s &= \mbox{Pr}\left( -1 + \frac{\sqrt{Y}}{X}
\mbox{Re}\left(U\right)
> 0 \right),\\
&= E \left \{ Q \left( \sqrt{\frac{2X^2}{Y}} \right) \right \},
\label{eq:mrc:main7}
\end{align}
\noindent where $Q\left(x\right)=\frac{1}{\sqrt{2\pi}}\int_x^\infty
e^{-\frac{t^2}{2}} dt$ is the Gaussian Q-function defined in
\cite{GradRzy00}. Defining $\gamma=X^2Y^{-1}$ gives the BPSK SER as
$E \left \{ Q \left( \sqrt{2\gamma} \right) \right \}$. Note that in
general $\gamma$ is a function of $\tilde{\bm{s}}$ but this
dependence is not shown for convenience. For BPSK, each element of
$\tilde{\bm{s}}$ has unit modulus and so there is no dependence on
$\tilde{\bm{s}}$ and the SER in \dref{eq:mrc:main7} is valid for any
values of $\tilde{\bm{s}}$. For many modulations \cite{Gold00,
Peter07}, SERs are constructed from similar functions of the form
\begin{align}
\mathcal{W}_1\left(a,b, \tilde{\bm{s}} \right) &= E \left \{ a Q
\left( \sqrt{b\gamma} \right) \right
\}, \label{eq:mrc:main88_Qfunction} \\
&= \int_0^\infty a Q \left( \sqrt{b\gamma} \right) f\left(\gamma
\right) d\gamma, \label{eq:mrc:main88}
\end{align}
\noindent where $f\left(\gamma\right)$ is the probability density
function (pdf) of $\gamma$. Hence, our approach involves averaging
the $Q$-function in \dref{eq:mrc:main88} over $\gamma$. There are
alternative routes to the same result. For example, the $Q$-function
in \dref{eq:mrc:main7} could be averaged over the joint distribution
of $X$ and $Y$. For some modulations, such as BPSK, the SER can be
given exactly in terms of $\mathcal{W}_1\left(a,b, \tilde{\bm{s}}
\right)$, whereas for other modulations it will provide an
approximation. Using integration by parts on \dref{eq:mrc:main88}
gives
\begin{equation}
\mathcal{W}_1\left(a,b, \tilde{\bm{s}} \right) = \frac{a}{\sqrt{2
\pi}} \int_0^\infty e^{-\frac{w^2}{2}} F_{\gamma}\left(\frac{
w^2}{b} \right) dw, \label{eq:mrc:ser1}
\end{equation}
\noindent where $F\left(.\right)$ is the cumulative distribution
function (cdf) of $\gamma$. Hence, SER performance for MRC relies on
the evaluation of \dref{eq:mrc:ser1} which in turn relies on the cdf
of $\gamma$. \\
In the microdiversity case, all the $\bm{P}_i$ matrices are
proportional to the identity and $\gamma$ reduces to a simplified
expression, $\gamma \propto \chi^2$, where $\chi^2$ is a chi-squared
random variable \cite{Miller64}. In the macrodiversity case, this
reduction does not occur and $\gamma$ is proportional, not to a
simple chi-squared random variable, but to a ratio of powers of
correlated quadratic forms. This is the novel analytical challenge
posed by the macrodiversity scenario.
The derivation of the cdf is based on the joint distribution of
$X,Y$. From \cite{Firag11}, the joint distribution of $X$ and $Y$
becomes
\begin{align}
\begin{split}
f_{X,Y}\left(x,y\right) &= \sum_{i=1}^{n_R} \sum_{k\neq i }^{n_R}
\xi_{ik} e^{-\frac{x}{P_{i1}}}e^{-\beta_{ik} \left( y
- \frac{Q_i x}{P_{i1}} \right)} \\
&\times \begin{cases} u\left(y - \frac{Q_i}{P_{i1}}x \right) u\left( x \right)  &  \mbox{for} \hspace{2mm} \beta_{ik}  > 0   \\
-u\left(\frac{Q_i}{P_{i1}}x - y\right)u\left( x \right) & \mbox{for}
\hspace{2mm} \beta_{ik} < 0,
\end{cases}
\end{split}
\label{eq:mrc:pdf4}
\end{align}
where $u\left(x \right)$ is the standard unit step function defined
as
\begin{align}
u \left( x \right) &= \begin{cases} 0 & x < 0 \\
                                    1 & x > 0,
\end{cases}
\label{eq:macro_mrc:unit_step_function}
\end{align}
\noindent and
\begin{align}
\xi_{ik} &= \frac{P_{i1}^{n_R-2} \left.\upsilon_{ik}\right.^{n_R
-3}}{\prod_{l\neq i,k}^{n_R} \left(\upsilon_{ik}\nu_{il} - \nu_{ik}\upsilon_{il}\right)}, \\
\beta_{ik} &= \frac{\nu_{ik}}{\upsilon_{ik}},\label{eq:mrc:beta} \\
\upsilon_{ik} &= P_{i1}Q_k - Q_i P_{k1}, \\
\nu_{ik} &= P_{i1}-P_{k1},\\
\bm{Q}&=\bm{D} \left( \tilde{\bm{s}} \right)\bm{P}_1
=\mbox{diag}\left(Q_1, Q_2, \dots, Q_{n_R}\right).
\end{align}
\noindent Note that each term in the summation of \dref{eq:mrc:pdf4}
has its own region of validity depending on the algebraic sign of
$\beta_{ik}$. For example, when $\beta_{ik} > 0$, the region of
validity becomes the infinite region bounded below by the $x=0$ and
$y = \frac{Q_i}{P_{i1}}x$ curves. The $\beta_{ik}=0$ condition has
been ignored since the case of distributed users with a single
antenna always yields $\beta_{ik}\neq0$.
The cdf of $\gamma$ is defined by
\begin{align}
F_{\gamma}\left( r \right)&= \mbox{Pr}\left( \gamma < r \right)=
\mbox{Pr}\left( \frac{X^2}{Y} < r \right), \\  &=\mbox{Pr}\left( X^2
-rY < 0 \right),\\
&=\iint_{\mathcal{D}} f_{X,Y}\left(x,y\right) dx dy,
\label{eq:mrc:cdf1}
\end{align}
\noindent where the domain of integration is defined by
${\mathcal{D}}= \left\{x,y: x\geq 0, y \geq 0 \hspace{2mm}
\mbox{and} \hspace{2mm} x^2 - ry < 0 \right\}$. In Appendix
\ref{sec:app3}, the integral in \dref{eq:mrc:cdf1} is computed
giving
\begin{align}
F_{\gamma}\left(r \right) &= \sum_{i=1}^{n_R} \sum_{k\neq i }^{n_R}
F_{ik}\left(r \right), \label{eq:mrc:cdf2}
\end{align}
\noindent where $F_{ik}\left(r \right)=F_{ik}^{1}\left(r \right)$
for $\beta_{ik}  > 0$ and $F_{ik}\left(r \right)=F_{ik}^{2}\left(r
\right)$ for $\beta_{ik}<0$, where $F_{ik}^{1}\left(r \right)$ and
$F_{ik}^{2}\left(r \right)$ are given in \dref{eq:mrc:app3:cdf12}
and \dref{eq:mrc:app3:cdf22} and $\beta_{ik}$ is given in
\dref{eq:mrc:beta}.
\begin{align}
\begin{split}
F_{ik}^{1}\left(r \right) &= \frac{P_{i1}\xi_{ik}}{\beta_{ik}}
\left( 1 - e^{-\frac{rQ_i}{P_{i1}^2}} \right) +
\frac{\xi_{ik}}{2\beta_{ik}} \sqrt{\frac{\pi r}{\beta_{ik}}}
e^{\frac{r\omega_{ik}^2}{4\beta_{ik}}} \\
& \qquad \qquad \qquad \qquad \times \left( 1 - \Phi \left( \sqrt{r
\beta_{ik}} \alpha_{ik} \right) \right),
\end{split}
\label{eq:mrc:app3:cdf12}
\end{align}
\begin{align}
\begin{split}
F_{ik}^{2}\left(r \right) &= \frac{P_{i1}\xi_{ik}}{\beta_{ik}}\left(
1 - e^{-\frac{rQ_i}{P_{i1}^2}} \right) -
\frac{\xi_{ik}}{2\beta_{ik}} \sqrt{\frac{\pi r}{-\beta_{ik}}}
e^{\frac{r\omega_{ik}^2}{4\beta_{ik}}} \\
& \times \left( \mbox{erfi}\left(\sqrt{-r \beta_{ik}} \alpha_{ik}
\right) + \mbox{erfi}\left( \frac{1}{2} \sqrt{\frac{r}{-\beta_{ik}}}
\omega_{ik} \right)\! \right),
\end{split}
\label{eq:mrc:app3:cdf22}
\end{align}
\noindent where
\begin{subequations} \label{eq:mrc:app3:omega_alpha1}
\begin{align}
\omega_{ik} &= \frac{1 - \beta_{ik}Q_i }{P_{i1}}, \quad \quad
\alpha_{ik} = \frac{Q_i}{P_{i1}} + \frac{\omega_{ik}}{2\beta_{ik}},
\end{align}
\end{subequations}
\noindent and $\Phi \left( x \right)=\frac{2}{\sqrt{\pi}}\int_0^x
e^{-t^2} dt$ in \dref{eq:mrc:GR_integral1} and
\dref{eq:mrc:GR_integral2} is the standard error function
\cite{GradRzy00}. Furthermore, $Q\left(x\right)=0.5 \left(1- \Phi
\left( \frac{x}{\sqrt{2}}\right)\right)$. For convenience, we expand
$\beta_{ik}$ in \dref{eq:mrc:beta} and also give the result here as
\begin{align}
\!\! \beta_{ik} \!=\!
\frac{\frac{1}{P_{i1}}-\frac{1}{P_{k1}}}{\sum_{u=2}^N \left( P_{ku}
- P_{iu} \right)\left|s_u\right|^2}.
\end{align}
Note that the case of $\beta_{ik}=0$ is not considered as this is
the case when $P_{i1}=P_{k1}$, an event which occurs with
probability zero when the receive antennas are not co-located. As
for the cdf, the SER analysis is performed separately according to
the algebraic sign of $\beta_{ik}$. Therefore, substituting
\dref{eq:mrc:cdf2} into \dref{eq:mrc:ser1}, the final result is
\begin{align}
\mathcal{W}_1 \left(a,b, \tilde{\bm{s}} \right) &= \sum_{i=1}^{n_R}
\sum_{k\neq i }^{n_R} P_{s_{ik}}, \label{eq:mrc:ser2}
\end{align}
\noindent where $P_{s_{ik}}=P_{s_{ik}}^{1}$ for $\beta_{ik}>0$ and
$P_{s_{ik}}=P_{s_{ik}}^{2}$ for $\beta_{ik}<0$, and $P_{s_{ik}}^{1}$
and $P_{s_{ik}}^{2}$ are given in \dref{eq:mrc:ser21} and
\dref{eq:mrc:ser22}.
\begin{table*}[t]\small
%\normalsize
\begin{align}
%\begin{split}
P_{s_{ik}}^{1} &= \frac{a}{\sqrt{2}} \left(
\frac{P_{i1}\xi_{ik}}{\beta_{ik}} \left( \frac{1}{\sqrt{2}} -
\frac{1}{2\sqrt{\left( \frac{1}{2} + \frac{Q_i}{bP_{i1}^2}\right)}}
\right) \right. +  \left. \sqrt{\frac{b}{\beta_{ik}}}
\frac{\xi_{ik}}{\left( \omega_{ik}^2 - 2b\beta_{ik} \right) } \left(
\sqrt{\frac{\beta_{ik}}{b \left( \frac{1}{2} + \frac{Q_i}{bP_{i1}^2}
\right)  }}\alpha_{ik}  - 1  \right) \right).
%\end{split}
\label{eq:mrc:ser21}
\end{align}

\begin{align}
%\begin{split}
P_{s_{ik}}^{2} &= \frac{a}{\sqrt{2 }} \left(
\frac{P_{i1}\xi_{ik}}{\beta_{ik}} \left( \frac{1}{\sqrt{2}} -
\frac{1}{2\sqrt{\left( \frac{1}{2} + \frac{Q_i}{bP_{i1}^2}\right)}}
\right) \right. +  \left. \sqrt{\frac{b}{-\beta_{ik}}}
\frac{\xi_{ik}}{\left( \omega_{ik}^2 - 2b\beta_{ik} \right) } \left(
\sqrt{\frac{-\beta_{ik}}{b \left( \frac{1}{2} +
\frac{Q_i}{bP_{i1}^2} \right)  }} \alpha_{ik}  +
\frac{\omega_{ik}}{\sqrt{-2b \beta_{ik}}} \right) \right).
%\end{split}
\label{eq:mrc:ser22}
\end{align}

\hrulefill \vspace{4pt}
%\end{figure*}
\end{table*}
The results in \dref{eq:mrc:ser21} and \dref{eq:mrc:ser22} are
obtained using the following three standard integral identities
\cite{GradRzy00}
%
%
%\begin{enumerate}
%%
%\item
\begin{align}
\int_0^\infty e^{\mu x}  \left( 1 -  \Phi \left( \sqrt{\alpha x
}\right) \right)& dx = \frac{1}{\mu} \left(
\sqrt{\frac{\alpha}{\alpha-\mu}} -1  \right), \nonumber \\
\mbox{for} \hspace{3mm} &\mbox{Re} \left( \alpha\right) > 0 ;
\mbox{Re} \left( \mu \right) < \mbox{Re} \left( \alpha\right),
\label{eq:mrc:GR_integral1}
\end{align}
%
%
%\item
\begin{align}
\int_0^\infty x e^{-\mu x^2}\Phi \left( j a x \right) dx &=
\frac{ja}{2\mu \sqrt{\mu - a^2} }, \nonumber \\
\mbox{for} \hspace{5mm} &\mbox{Re}\left( \mu \right) > 0;
\mbox{Re}\left( \mu \right)
> \mbox{Re}\left( a^2 \right), \label{eq:mrc:GR_integral2}
\end{align}
%
%\item
\begin{align}
\int_0^\infty e^{-q^2 x^2}dx &= \frac{\sqrt{\pi}}{2q}  \hspace{5mm}
\mbox{for} \hspace{5mm} q > 0.
\end{align}
For multi-level constellations, the values of $\tilde{\bm{s}}$
affect $\bm{D}\left(\tilde{\bm{s}}\right)$ and therefore $\gamma$.
Hence, SER results must average \dref{eq:mrc:ser2} over all possible
values of $\tilde{\bm{s}}$. This gives
\begin{align}\label{eq:mrc:general_ser1}
\mathcal{W}_1 \left(a,b \right) &= \sum_{\tilde{\bm{s}}}
\mathcal{W}_1 \left(a,b, \tilde{\bm{s}} \right) \mbox{Pr}\left(
\tilde{\bm{s}} \right),
\end{align}
\noindent where \dref{eq:mrc:general_ser1} may be an exact or
approximate SER result, the summation is over all possible
$\tilde{\bm{s}}$ and $\mbox{Pr}\left( \tilde{\bm{s}} \right)$ is the
probability of a particular $\tilde{\bm{s}}$ value. Finally, for
BPSK modulation, the SER in \dref{eq:mrc:main7} becomes
\begin{align}
P_s &= \mathcal{W}_1 \left(1,2, \tilde{\bm{s}} \right).
\end{align}
%
%
%\end{enumerate}
%
%
\subsection{Extended SER Analysis}\label{sub_sec:extended_SER_analysis}
For $M$-QAM, first order SER approximations can be found via
expressions of the form in \dref{eq:mrc:main88_Qfunction}. Exact
results involve expectation over the $Q^2(.)$ function in addition
to \dref{eq:mrc:main88_Qfunction}. As an example, consider 4-QAM
where the SER is given by
\begin{align}
\begin{split}
P_s = \mbox{Pr} &\left(-\frac{1}{\sqrt{2}} +
\frac{1}{\sqrt{\gamma}}\mbox{Re}\left(U\right) > 0 \hspace{2mm}
\right. \\
& \qquad \qquad \mbox{or} \left. \hspace{2mm} -\frac{1}{\sqrt{2}} +
\frac{1}{\sqrt{\gamma}}\mbox{Im}\left(U\right) > 0 \right)
\end{split}
\end{align}
\begin{align}
P_s &= 1 - \mbox{Pr}\left( \mbox{Re}\left( U \right) <
\sqrt{\frac{\gamma}{2}}
\right)^2 \\
&= 1 - E\left \{ \left( 1 - Q\left( \sqrt{\gamma} \right)
\right)^2 \right \} \\
&= 2 E\left \{ Q\left( \sqrt{\gamma} \right) \right \} -  E\left \{
Q^2\left( \sqrt{\gamma} \right) \right \}. \label{eq:mrc:ser_QPSK1}
\end{align}
Here, the $2 E\left \{ Q\left( \sqrt{\gamma} \right) \right
\}=\mathcal{W}_1 \left(2,1, \tilde{\bm{s}} \right)$ term in
\dref{eq:mrc:ser_QPSK1} is a good approximation to $P_s$
\cite{Proakis01} and the remaining term, $E\left \{ Q^2\left(
\sqrt{\gamma} \right) \right \}$, makes only a small adjustment.
However, in other variations of $M$-QAM modulation schemes the
contribution from $Q^2\left(.\right)$ is not negligible
\cite{Proakis01}. Therefore, for general $M$-QAM, the exact SER is
useful and this can be written in terms of $\mathcal{W}_1 \left(a,b,
\tilde{\bm{s}} \right)$ and $\mathcal{W}_2 \left(a,b, \tilde{\bm{s}}
\right)=E\left \{a Q^2\left( \sqrt{b\gamma} \right) \right \}$. The
first expectation is found in \dref{eq:mrc:ser2}. The second
expectation can be derived as follows. Let,
\begin{align}
\mathcal{W}_2 \left(a,b, \tilde{\bm{s}} \right) &= a \int_0^\infty
Q^2 \left( \sqrt{b\gamma} \right) f\left(\gamma \right) d\gamma.
\label{eq:macro_mrc:Q2_expectation}
\end{align}
Using integration by parts on \dref{eq:macro_mrc:Q2_expectation}
gives
\begin{align}
\!\!\!\mathcal{W}_2 \left(a,b, \tilde{\bm{s}} \right) &= a
\sqrt{\frac{2}{\pi}} \int_0^\infty e^{-\frac{w^2}{2}}Q\left(w
\right) F_{\gamma}\left(\frac{w^2}{b} \right) dw.
\label{eq:macro_mrc:Q2_expectation1}
\end{align}
In order to facilitate our analysis we need two fundamental
probability integrals. Therefore, we derive both integrals in
Appendix \ref{sec:app4} along with their regions of convergence,
since they may have applications in other communication problems.
Note that similar results may be found in \cite{Prudnikov86}, but
these are for restricted ranges of the parameter values. The
macrodiversity integrals require a wider range of values and the
analysis in Appendix \ref{sec:app4} enables us to evaluate both the
integral values and the precise region of validity.
As for the simple SER analysis, the extended analysis is also
performed separately according to the algebraic sign of
$\beta_{ik}$. Therefore, substituting \dref{eq:mrc:cdf2} into
\dref{eq:macro_mrc:Q2_expectation1}, the final result is derived in
Appendix \ref{sec:app4} as
\begin{align}
\mathcal{W}_2 \left(a,b, \tilde{\bm{s}} \right) &= \sum_{i=1}^{n_R}
\sum_{k\neq i }^{n_R} \tilde{P}_{s_{ik}},
\label{eq:mrc:expedted_Q22}
\end{align}
\noindent where $\tilde{P}_{s_{ik}}=\tilde{P}_{s_{ik}}^{1}$  for
$\beta_{ik}>0$ and $\tilde{P}_{s_{ik}}=\tilde{P}_{s_{ik}}^{2}$ for
$\beta_{ik}<0$, where $\tilde{P}_{s_{ik}}^{1}$ and
$\tilde{P}_{s_{ik}}^{2}$ are given in \dref{eq:mrc:Q2_ser1} and
\dref{eq:mrc:Q2_ser2}, respectively.
\begin{table*}[t]\small
\begin{align}
%\begin{split}
\tilde{P}_{s_{ik}}^{1} &= \frac{a}{\sqrt{2}} \left(
\frac{P_{i1}\xi_{ik}}{\beta_{ik}} \left(  \frac{1}{\sqrt{8}} -
\frac{\tan^{-1}\left( \sqrt{1 + \frac{2 Q_i}{bP_{i1}^2}}
\right)}{\pi \sqrt{ \left( \frac{1}{2} +
\frac{Q_i}{bP_{i1}^2}\right)}} \right) + \frac{\xi_{ik}}{\sqrt{b
\beta_{ik}} \beta_{ik}} I_2 \left(
\alpha_{ik}\sqrt{\frac{\beta_{ik}}{b}}, \left.
\frac{\omega_{ik}^2}{4b\beta_{ik}} - \frac{1}{2}\right. \right)
\right). \label{eq:mrc:Q2_ser1}
\end{align}
\begin{align}
\begin{split}
\tilde{P}_{s_{ik}}^{2} \!=\! \frac{a}{\sqrt{2}}\! \left(\!
\frac{P_{i1}\xi_{ik}}{\beta_{ik}} \!\left(  \frac{1}{\sqrt{8}} \!-
\! \frac{\tan^{-1}\! \! \left( \sqrt{1 + \frac{2 Q_i}{bP_{i1}^2}}
\right)}{\pi \sqrt{ \left( \frac{1}{2} \!+ \!
\frac{Q_i}{bP_{i1}^2}\right)}} \right) \right. \! \! &-
\frac{1}{j}\frac{\xi_{ik}}{\sqrt{-b \beta_{ik}} \beta_{ik}}  \\
 \! \! & \times \!\! \left. \left( I_1 \! \! \left(
\alpha_{ik}\sqrt{\frac{-\beta_{ik}}{b}}, \left. \frac{1}{2} -
\frac{\omega_{ik}^2}{4b\beta_{ik}} \right. \right) \! + \! I_1 \! \!
\left(\frac{\omega_{ik}}{2\sqrt{-b \beta_{ik}}}, \left. \frac{1}{2}
- \frac{\omega_{ik}^2}{4b\beta_{ik}} \right. \right) \right)
\right). \label{eq:mrc:Q2_ser2}
\end{split}
\end{align}
\hrulefill \vspace{4pt}
%\end{figure*}
\end{table*}
Hence, the exact SERs are computable using \dref{eq:mrc:ser2} and
\dref{eq:mrc:expedted_Q22} for any $M$-QAM modulation. As in Sec.
\ref{sub_sec:simple_SER_analysis}, for multi-level constellations
the SER results depend on $\mathcal{W}_1 \left(a,b\right)$ and
$\mathcal{W}_2 \left(a,b\right)$ results where $\mathcal{W}_1
\left(a,b\right)$ is given in \dref{eq:mrc:general_ser1} and
\begin{align}\label{eq:mrc:general_ser2}
\mathcal{W}_2 \left(a,b \right) &= \sum_{\tilde{\bm{s}}}
\mathcal{W}_2 \left(a,b, \tilde{\bm{s}} \right) \mbox{Pr}\left(
\tilde{\bm{s}} \right).
\end{align}
For QPSK modulation the SER in \dref{eq:mrc:ser_QPSK1} becomes
\begin{align}
    P_s &= \mathcal{W}_1 \left(2,1, \tilde{\bm{s}}\right) - \mathcal{W}_2 \left(1,1,
    \tilde{\bm{s}}\right).
\end{align}

\subsection{A Simple Power Metric}\label{sub_sec:power_metric}
The SER and any other performance metrics are functions of the power
matrices $\bm{P}_1, \bm{P}_2, \dots, \bm{P}_N$. Although
\dref{eq:mrc:ser2} and \dref{eq:mrc:expedted_Q22} give the exact SER
as a function of these powers, the result is complex and does not offer any simple insights into the relationship between
performance and the powers. Hence, we consider \dref{eq:mrc:main3}
and \dref{eq:mrc:z} which give the mean SINR of the combiner as
\begin{align}
\tilde{m}_P &\triangleq  E \left\{ \frac{\left( \bm{h}_1^H \bm{h}_1
\right)^2}{\bm{h}_1^H \left(\sum_{k=2}^N \bm{h}_k \bm{h}_k^H +
\sigma^2 \bm{I} \right)\bm{h}_1} \right\},
\label{eq:mrc:simple_metric}
\end{align}
where $\tilde{m}_P$ is a performance metric based on the link powers
and we have used $E\left \{ \left |s_i \right|^2 \right\}=1$. Exact
evaluation of \dref{eq:mrc:simple_metric} is possible but it is
rather involved and produces complex expressions. Hence, we prefer
the compact approximation based on the first order delta method,
similar to the Laplace approximation \cite{Lib94}, given by
\begin{align}
m_P &=   \frac{ E \left\{ \left( \bm{h}_1^H \bm{h}_1 \right)^2
\right\} }{ E \left\{ \bm{h}_1^H \left(\sum_{k=2}^N \bm{h}_k
\bm{h}_k^H + \sigma^2 \bm{I} \right)\bm{h}_1 \right\} }.
\label{eq:mrc:simple_metric2}
\end{align}
Using established results for the moments of quadratic forms
\cite[pp.~119]{Miller64}, we obtain
\begin{align}
m_P &=   \frac{\Tr\left( \bm{P}_1\right)^2 +
\Tr\left(\bm{P}_1^2\right)}{\Tr \left (\sum_{i=2}^N \bm{P}_1\bm{P}_i
+ \sigma^2 \bm{P}_1 \right)}. \label{eq:mrc:simple_metric3}
\end{align}
From $\dref{eq:mrc:simple_metric3}$, we observe that while a
$\bm{P}_1$ matrix with large entries boosts the numerator, hence
improving performance, it also interacts with the interferers in the
denominator. Since MRC is based on weighting the strongest signal,
the most advantageous interference profile is for the stronger
interferers to line up with the weaker desired signals and
vice-versa. This intuitive result is precisely captured by
\dref{eq:mrc:simple_metric3} which increases with $\Tr\left(
\bm{P}_1\right)$ and also increases as $\Tr \left (\bm{P}_1\bm{P}_i
\right)$ decreases, for $i=2,3, \dots, N$.\\
Although $m_P$ captures some of the important relationships between
performance and the power matrices, it is not always an accurate
predictor of performance. As the SINR grows, the mean becomes
further from the lower tail which governs error rate performance.
Hence, we expect the mean to carry less information about SER at
high SINR. This is discussed in more detail in Sec.
\ref{sec:numerical_analysis}.
\subsection{Remarks on Systems with Multiple Co-located Receive Antennas at TPs}\label{sec:remarks}
The analysis in Sec. \ref{sec:perform_analysis} is restricted to
situations where the TPs have a single antenna each. However, if the
receiver, for example has two co-located antennas at any TP, the
system analysis still can be handled by the same method, but will
result in a different joint distribution for \dref{eq:mrc:pdf4}.
Thus, every new scenario for co-located antennas gives a new joint
distribution and in turn this gives a different error rate
expression.
A pragmatic solution is to use a perturbation approach. If
$P_{i1}=P_{r1}$ (corresponding to receive antennas $i$ and $r$ of
the desired user, being co-located at $i$\textsuperscript{th} TP)
then we can use $P_{i1}$ and $P_{i1}+\epsilon$ for the two powers
where $\epsilon$ is a small perturbation. This approach provides
stable and accurate results as will be shown in Sec.
\ref{sec:numerical_analysis}.
\section{Numerical and Simulation Results}\label{sec:numerical_analysis}
\begin{figure}[h]
\centerline{\includegraphics*[scale=0.65]{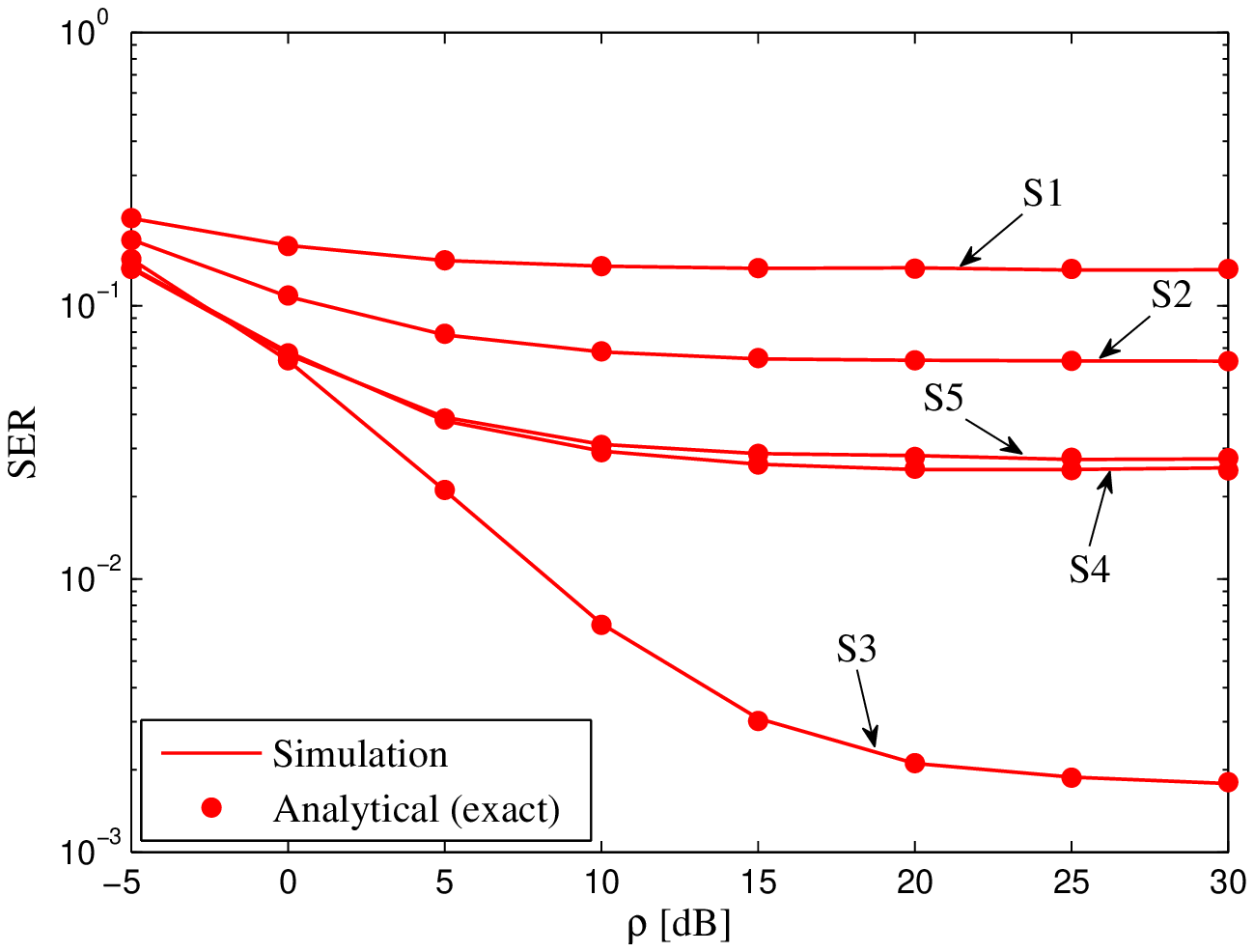}}
\caption{Analytical and simulated SER values for a MRC receiver with
BPSK modulation in flat Rayleigh fading for scenarios S1-S5 with
parameters: $n_R=3$ and $\varsigma=1$.} \label{fig:macro_mrc:2}
\end{figure}

\begin{figure}[h]
\centerline{\includegraphics*[scale=0.65]{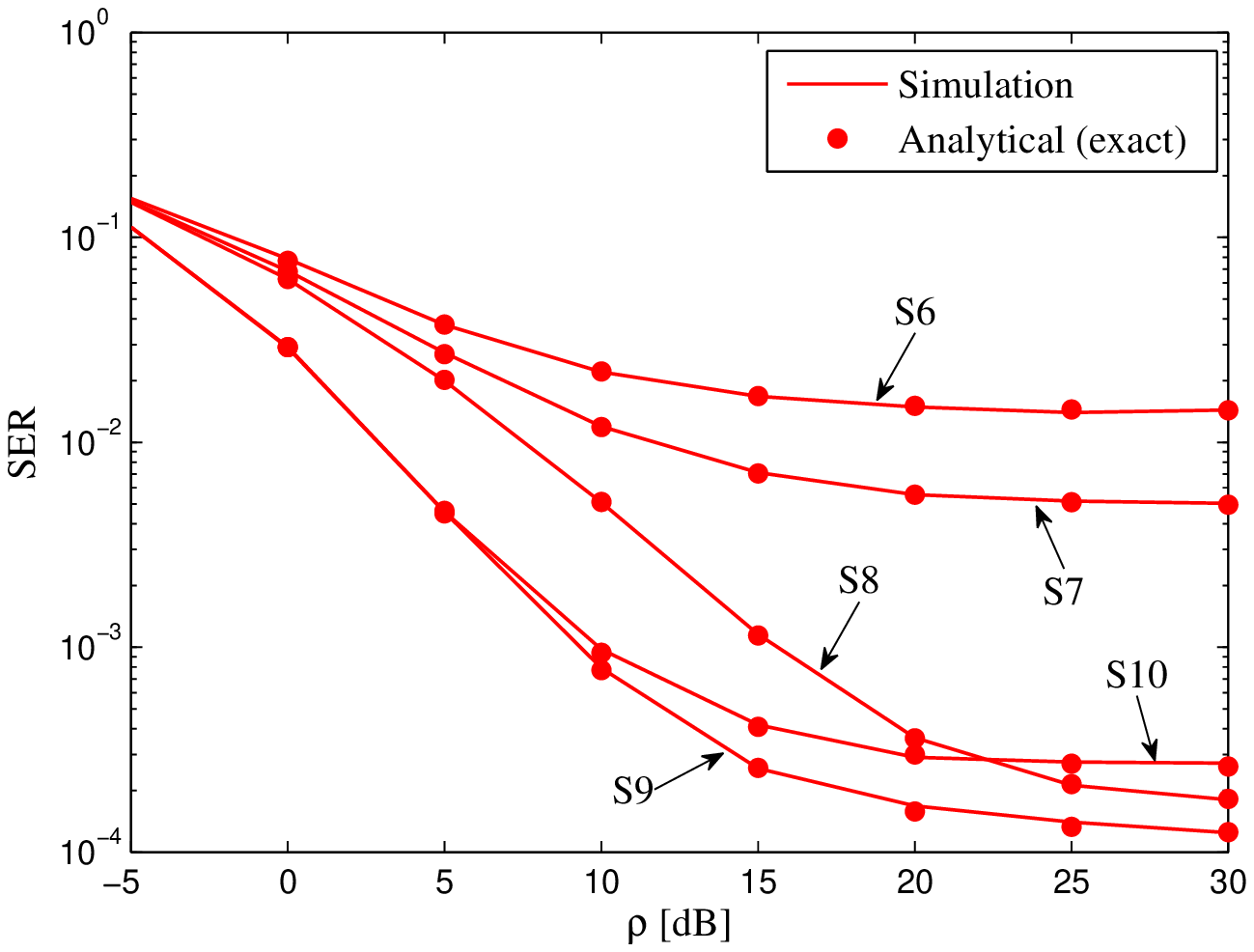}}
\caption{Analytical and simulated SER values for a MRC receiver with
BPSK modulation in flat Rayleigh fading for scenarios S6-S10 with
parameters: $n_R=3$ and $\varsigma=10$.} \label{fig:macro_mrc:3}
\end{figure}

For the numerical results, we consider a system with three
distributed receive antennas and also a larger system with 6 receive
antennas deployed in three sets of co-located pairs. Hence, there
are three positions at which one or two antennas are deployed and
these are refereed to as locations. Note that the number of
interferers in the system is irrelevant, since, from
\dref{eq:mrc:main5}, their effect is governed by $\sum_{k=2}^{N}
\bm{P}_k \left| s_k \right|^2$. Hence, one interferer with a power
matrix equal to $\sum_{k=2}^{N} \bm{P}_k \left| s_k \right|^2$ is
equivalent to $N-1$ interferers with power matrices $\bm{P}_2 \left|
s_2 \right|^2, \dots, \bm{P}_N \left| s_N \right|^2$. Hence, we
consider a single interferer throughout. In this section we consider
BPSK and 4-QAM results where $|s_i|^2=1$ $\forall i$. Hence, for
both systems, we parameterize the performance by three parameters
which are independent of the transmit symbols. The average received
signal to noise ratio is defined by $\rho= \left. \Tr\left(
\bm{P}_1\right) \right. / n_R \sigma^2$. The total signal to
interference ratio is defined by $\varsigma = \Tr\left(
\bm{P}_1\right) / \Tr\left( \bm{P}_2\right)$. The spread of the
signal power across the three locations is assumed to follow an
exponential profile, as in \cite{Gao98}, so that a range of
possibilities can be covered with only one parameter. The
exponential profile is defined by
\begin{align}
P_{ik} &= K_k \left( \alpha \right) \alpha^{i-1},
\end{align}
for receive location i and source k where
\begin{align}
K_k \left( \alpha \right) &= \Tr\left( \bm{P}_k\right) /
\left(1+\alpha + \alpha^2 \right), \quad k=1,2,
\end{align}
and $\alpha > 0$ is the parameter controlling the uniformity of the
powers across the antennas. Note that as $\alpha \rightarrow 0$ the
received power is dominant at the first location, as $\alpha$
becomes large $\left( \alpha \gg 1 \right)$ the third location is
dominant and as $\alpha \rightarrow 1$ there is an even spread, as
in the standard microdiversity scenario. In Figs.
\ref{fig:macro_mrc:2}-\ref{fig:macro_mrc:3} we show SER results for
the ten scenarios (S1-S10) given in Table \ref{table:macro_mrc:1}.
\begin{table}[Parameters for Figures]
    \caption{Parameters for Figures \ref{fig:macro_mrc:2} and \ref{fig:macro_mrc:3}}
    \centering
    \begin{tabular}{ r | c | l | l | c | l }
    \hline \hline
    & & \multicolumn{2}{|c|}{Decay Parameter} & &\\ \cline{3-4}
    Sc. No. & $\varsigma$ & Desired & Interfering & $\overset{m_P}{\left(\mbox{dB}\right)}$ & Err. Floor \\ \hline
    S1 & 1 & $\alpha=\frac{1}{65}$ & $\alpha=\frac{1}{65}$ & 3.06 & 1.36e-1 \\
    S2 & 1 & $\alpha=\frac{1}{65}$ & $\alpha=1$ & 7.68 & 6.26e-2 \\
    S3 & 1 & $\alpha=\frac{1}{65}$ & $\alpha=65$ & 28.64 & 1.80e-3\\
    S4 & 1 & $\alpha=1$ & $\alpha=1$ & 5.97 & 2.49e-2\\
    S5 & 1 & $\alpha=1$ & $\alpha=\frac{1}{65}$ & 5.97 &  2.76e-2 \\ \hline
    S6 & 10 & $\alpha=\frac{1}{65}$ & $\alpha=\frac{1}{65}$ & 12.93 & 1.42e-2 \\
    S7 & 10 & $\alpha=\frac{1}{65}$ & $\alpha=1$ & 17.30 & 4.90e-3\\
    S8 & 10 & $\alpha=\frac{1}{65}$ & $\alpha=65$ & 27.62 & 1.68e-4\\
    S9 & 10 & $\alpha=1$ & $\alpha=1$ & 15.60 & 1.21e-4\\
    S10 & 10 & $\alpha=1$ & $\alpha=\frac{1}{65}$ & 15.60 & 2.57e-4\\
    \hline
    \end{tabular}
\label{table:macro_mrc:1} % is used to refer this table in the text
\end{table}
\begin{table}[Parameters for Figures]
    \caption{Parameters for Figure \ref{fig:macro_mrc:4}}
    \centering
    \begin{tabular}{ r | c | l | l | c | l }
    \hline \hline
    & & \multicolumn{2}{|c|}{Decay Parameter} & &\\ \cline{3-4}
    Sc. No. & $\varsigma$ & Desired & Interfering & $\overset{m_P}{\left(\mbox{dB}\right)}$ & Err. Floor \\ \hline
    S11 & 30 & $\alpha=\frac{1}{65}$ & $\alpha=\frac{1}{65}$ & 17.42& 1.54e-2\\
    S12 & 30 & $\alpha=\frac{1}{65}$ & $\alpha=1$ & 21.34 & 5.20e-3\\
    S13 & 30 & $\alpha=\frac{1}{65}$ & $\alpha=65$ & 27.68 & 1.99e-4\\
    S14 & 30 & $\alpha=1$ & $\alpha=1$ & 19.65 & 7.68e-5\\
    S15 & 30 & $\alpha=1$ & $\alpha=\frac{1}{65}$ & 19.64 & 1.72e-4\\
    \hline
    \end{tabular}
\label{table:macro_mrc:2} % is used to refer this table in the text
\end{table}
Note that an error floor occurs as $\rho \rightarrow \infty$ $\left(
\sigma^2 \rightarrow 0 \right)$ for fixed $\varsigma$. The value of
the error floor is obtained by letting $\sigma^2 \rightarrow 0$ in
\dref{eq:mrc:ser2}. In Table \ref{table:macro_mrc:1} we report the
values of $m_P$ and the error floor for the ten scenarios
considered. Note that $m_p$ is given for a $\sigma^2$ value
corresponding to $\rho=20$ dB.

\begin{table}[Parameters for Figures]
    \caption{Parameters for Figure \ref{fig:macro_mrc:5}}
    \centering
    \begin{tabular}{ r | c | l | l | c | l }
    \hline \hline
    & & \multicolumn{2}{|c|}{Decay Parameter} & &\\ \cline{3-4}
    Sc. No. & $\varsigma$ & Desired & Interfering & $\overset{m_P}{\left(\mbox{dB}\right)}$ & Err. Floor \\ \hline
    S16 & 20 & $\alpha=\frac{1}{65}$ & $\alpha=\frac{1}{65}$ & 17.57 & 1.50e-3\\
    S17 & 20 & $\alpha=\frac{1}{65}$ & $\alpha=1$ & 21.69 & 1.61e-4\\
    S18 & 20 & $\alpha=\frac{1}{65}$ & $\alpha=65$ & 29.32 & 5.51e-7\\
    S19 & 20 & $\alpha=1$ & $\alpha=1$ & 20.57 & 1.54e-7 \\
    S20 & 20 & $\alpha=1$ & $\alpha=\frac{1}{65}$ & 19.96 & 1.04e-6\\
    \hline
    \end{tabular}
\label{table:macro_mrc:3} % is used to refer this table in the text
\end{table}

\begin{figure}[h]
\centerline{\includegraphics*[scale=0.65]{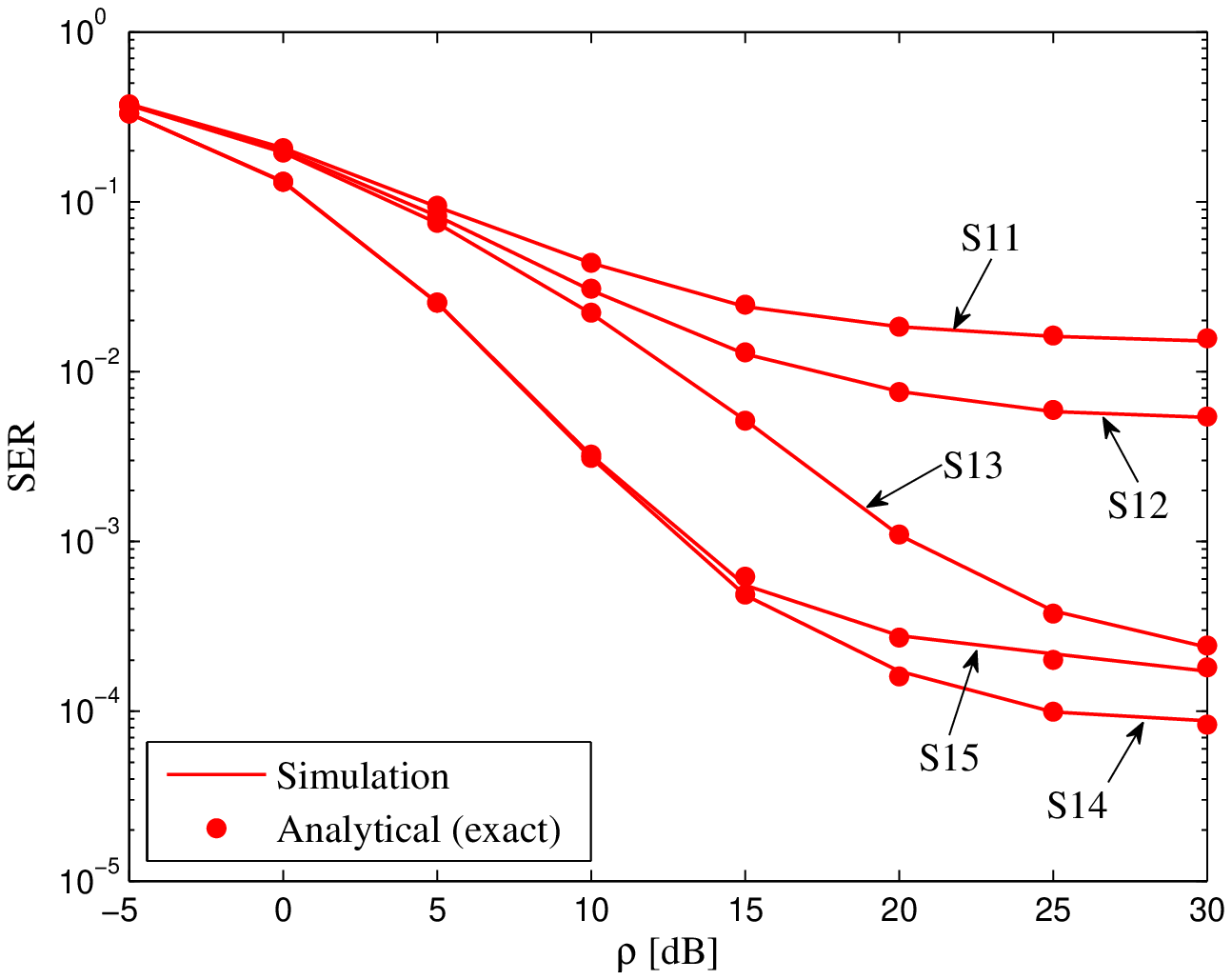}}
\caption{Analytical and simulated SER values for a MRC receiver with
QPSK modulation in flat Rayleigh fading for scenarios S11-S15 with
parameters: $n_R=3$ and $\varsigma=30$.} \label{fig:macro_mrc:4}
\end{figure}
\begin{figure}[h]
\centerline{\includegraphics*[scale=0.65]{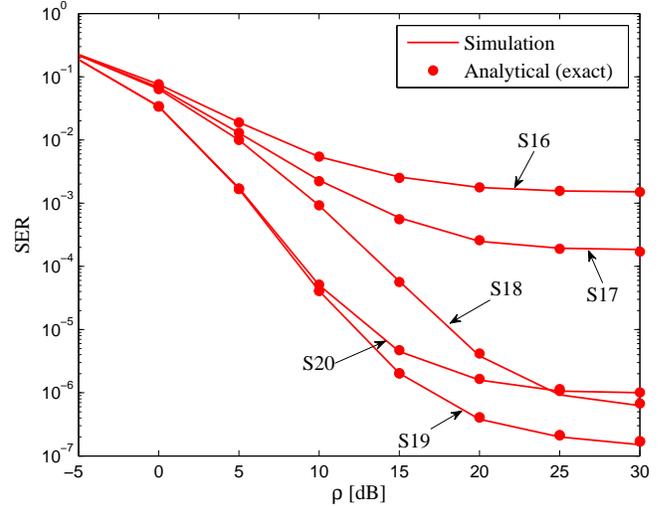}}
\caption{Analytical and simulated SER values for a MRC receiver with
QPSK modulation in flat Rayleigh fading for scenarios S16-S20 with
parameters: $n_R=6$ and $\varsigma=20$.} \label{fig:macro_mrc:5}
\end{figure}
Figures \ref{fig:macro_mrc:2} and \ref{fig:macro_mrc:3} verify the
analytical results in \dref{eq:mrc:ser2} for BPSK modulation with
simulations and also explore the effect of different power profiles.
In Fig. \ref{fig:macro_mrc:2}, a low SIR is considered with
$\varsigma=1$. Here, S1 is the worst case since the desired signal
profile is aligned with the interferer and the profile is rapidly
decaying giving little diversity. S3 is the best since the profiles
are opposing and the best desired signal aligns with the weakest
interference. Since Fig. \ref{fig:macro_mrc:2} has a low SIR the
major impact on performance is caused by the presence or absence of
a high SIR or low SIR at each antenna. In Fig.
\ref{fig:macro_mrc:3}, the same power profiles are considered but at
higher SIR, $\varsigma=10$, the order is changed. S6 is still the
worst as this scenario has high interference at all antennas and
little diversity. In contrast S8 is no longer the best with S9 now
giving better performance. Note that S9 has greater diversity with
an even spread of power across the antennas
and this becomes more important at high SIR.\\
Another comparison between scenarios can be seen in Table
\ref{table:macro_mrc:1}. Note that in Fig. \ref{fig:macro_mrc:2} the
ordering based on $m_P$ correctly identifies the best and worst
scenarios whereas in Fig. \ref{fig:macro_mrc:3} the $m_p$ metric
suggest that S8 is best whereas S9 is better. The $m_p$ metric gives
some intuition about macrodiversity MRC performance, especially at
low SIR, but it doesn't accurately capture diversity effects (seen
in the lower tail of the combiner output) which are
needed for accurate performance prediction.\\
In Fig. \ref{fig:macro_mrc:4}, the same power profiles are
considered for QPSK transmission. Here, the exact results from Sec.
\ref{sub_sec:extended_SER_analysis} are verified by simulation. In
particular, the SER expression in \dref{eq:mrc:ser_QPSK1} for QPSK
modulation is used along with \dref{eq:mrc:ser2} and
\dref{eq:mrc:expedted_Q22}. The relative performance provided by the
5 scenarios is the same as in Fig. \ref{fig:macro_mrc:3} except that
the cross over of S13 and S15 in Fig. \ref{fig:macro_mrc:4}
(equivalent to the cross over
of S8 and S10 in Fig. \ref{fig:macro_mrc:3}) does not occur until $\mbox{SNR} > 30$ dB.\\
Finally, in Fig. \ref{fig:macro_mrc:5} we consider the six antenna
receiver where antennas 1,2 are co-located, antennas 3,4 are
co-located elsewhere and antennas 5,6 are also co-located and
separated from antennas 1-4. Here, the long term receive SNR of a
source at antennas 1 and 2 will be the same. Hence, we use the
perturbation approach of Sec. \ref{sec:remarks} to obtain results.
Fig. \ref{fig:macro_mrc:5} validates the perturbation approach by
simulation and shows a large performance improvement relative to
Fig. \ref{fig:macro_mrc:4} due to the increased number of antennas.
Again, the results due to the five scenarios follow the same order
as in Figs. \ref{fig:macro_mrc:3} and \ref{fig:macro_mrc:4}. Note
that when $\alpha=1$ for both desired and interfering sources, the
system layout is microdiversity. Hence, scenarios S4, S9, S14 and
S19 provide microdiversity results.
\section{Conclusion}\label{sec:conclusion}
Exact SER results are derived for BPSK and $M$-QAM modulations in a
Rayleigh fading macrodiversity system employing MRC. The results
have applications to several systems of current interest in
communications including network MIMO and cooperative
communications. The analysis is used to study the effects of the
macrodiversity power profiles on MRC performance. It is shown that
simple power metrics may capture several features of MRC performance
but the impact of diversity in a distributed system is important at
realistic SINR values. Here, the exact results are necessary to
provide an accurate performance measure. In general, performance
improves as the desired signal dominates the interferer at some
antennas and as the desired power is spread more evenly over the
receive antennas. The exact balance between these two key features
is difficult to obtain in a simple form but is provided by the exact
solutions given.

\appendices
\section{Calculation of the cdf of $\gamma$ }\label{sec:app3}
Since each term in the summation of \dref{eq:mrc:pdf4} depends on
the algebraic sign of $\beta_{ik}$, the final cdf has two parts as
below
\begin{align}
F_{\gamma}\left(r \right) &= \sum_{i=1}^{n_R} \sum_{k\neq i }^{n_R}
F_{ik}\left(r \right), \label{eq:mrc:app3:cdf0}
\end{align}
\noindent where $F_{ik}\left(r \right)=F_{ik}^{1}\left(r \right)$
for $\beta_{ik}  > 0$ and $F_{ik}\left(r \right)=F_{ik}^{2}\left(r
\right)$ for $\beta_{ik}<0$. In subsection \ref{subsec:3A} we derive
$F_{ik}^{1}\left(r \right)$ followed by the derivation of
$F_{ik}^{2}\left(r \right)$ in subsection \ref{subsec:3B}.
\subsection{Derivation of $F_{ik}^{1}\left(r \right)$} \label{subsec:3A}
From the joint pdf in \dref{eq:mrc:pdf4}, when $\beta_{ik}  > 0$,
$F_{ik}^{1}\left(r \right)$ is given by
\begin{align}
F_{ik}^{1}\left(r \right) &= \iint_{\mathcal{F}_1}
f_{X,Y}\left(x,y\right) dx dy,
\end{align}
\noindent where $\mathcal{F}_1 = \left \{ x,y: x,y \geq 0, y -
\frac{Q_i}{P_{i1}}x \geq 0, y - \frac{x^2}{r} \geq 0 \right \}$. By
using standard methods for 2-D integrals we arrive at
\begin{align}
F_{ik}^{1}\left(r \right) &= \frac{P_{i1}\xi_{ik}}{\beta_{ik}} -
\xi_{ik}\int_{\frac{rQ_i}{P_{i1}}}^\infty
\int_{\frac{xQ_i}{P_{i1}}}^{\frac{x^2}{r}} e^{-\frac{x}{P_{i1}}}
e^{-\beta_{ik} \left( y - \frac{Q_i x}{P_{i1}} \right)} dy dx.
\label{eq:mrc:app3:cdf11}
\end{align}
The final result then becomes
\begin{align}
\begin{split}
F_{ik}^{1}\left(r \right) &= \frac{P_{i1}\xi_{ik}}{\beta_{ik}}
\left( 1 - e^{-\frac{rQ_i}{P_{i1}^2}} \right) +
\frac{\xi_{ik}}{2\beta_{ik}} \sqrt{\frac{\pi r}{\beta_{ik}}}
e^{\frac{r\omega_{ik}^2}{4\beta_{ik}}} \\
& \qquad \qquad \qquad \qquad \times \left( 1 - \Phi \left( \sqrt{r
\beta_{ik}} \alpha_{ik} \right) \right),
\end{split}
\label{eq:mrc:app3:cdf12_app}
\end{align}
\noindent where
\begin{subequations} \label{eq:mrc:app3:omega_alpha1_app}
\begin{align}
\omega_{ik} &= \frac{1 - \beta_{ik}Q_i }{P_{i1}}, \\
\alpha_{ik} &= \frac{Q_i}{P_{i1}} + \frac{\omega_{ik}}{2\beta_{ik}}.
\end{align}
\end{subequations}

The expression in \dref{eq:mrc:app3:cdf12} follows using standard
methods of integration in \dref{eq:mrc:app3:cdf11} and employing the
following integral identity \cite{GradRzy00} where necessary:
\begin{align}
\int_\alpha^\infty e^{-\beta x^2} dx &=
\frac{1}{2}\sqrt{\frac{\pi}{\beta}} \left( 1 - \Phi \left(
\sqrt{\beta} \alpha \right) \right).
\end{align}
\subsection{Derivation of $F_{ik}^{2}\left(r \right)$} \label{subsec:3B}
From the joint pdf in \dref{eq:mrc:pdf4}, when $\beta_{ik}  <  0$,
$F_{ik}^{2}\left(r \right)$ is given by
\begin{align}
F_{ik}^{2}\left(r \right) &= \iint_{\mathcal{F}_2}
f_{X,Y}\left(x,y\right) dx dy,
\end{align}
\noindent where $\mathcal{F}_2 = \left \{ x,y: x,y \geq 0, y -
\frac{Q_i}{P_{i1}}x \leq 0, y - \frac{x^2}{r} \geq 0 \right \}$. By
using standard methods for 2-D integrals we arrive at
\begin{align}
F_{ik}^{2}\left(r \right) &= - \xi_{ik}\int_0^{\frac{rQ_i}{P_{i1}}}
\int_{\frac{x^2}{r}}^{\frac{xQ_i}{P_{i1}}} e^{-\frac{x}{P_{i1}}}
e^{-\beta_{ik} \left( y - \frac{Q_i x}{P_{i1}} \right)} dy dx.
\label{eq:mrc:app3:cdf21}
\end{align}
The final result then becomes
\begin{align}
\begin{split}
F_{ik}^{2}\left(r \right) &= \frac{P_{i1}\xi_{ik}}{\beta_{ik}}\left(
1 - e^{-\frac{rQ_i}{P_{i1}^2}} \right) -
\frac{\xi_{ik}}{2\beta_{ik}} \sqrt{\frac{\pi r}{-\beta_{ik}}}
e^{\frac{r\omega_{ik}^2}{4\beta_{ik}}} \\
& \times \left( \mbox{erfi}\left(\sqrt{-r \beta_{ik}} \alpha_{ik}
\right) + \mbox{erfi}\left( \frac{1}{2} \sqrt{\frac{r}{-\beta_{ik}}}
\omega_{ik} \right)\! \right),
\end{split}
\label{eq:mrc:app3:cdf22_app}
\end{align}
\noindent where
\begin{align}
\mbox{erfi}\left(x \right) &= \frac{\Phi\left(jx\right)}{j}.
\end{align}

The function $\mbox{erfi}\left(.\right)$ is the error function with
a complex argument defined in \cite{GradRzy00}. Note that the square
roots appearing in \dref{eq:mrc:app3:cdf22} are the positive square
root of $\beta_{ik}$. The expression in \dref{eq:mrc:app3:cdf22}
follows using standard methods of integration and employing the
following integral identity \cite{GradRzy00} where necessary:
\begin{align}
\int e^{a x^2} dx &= \frac{1}{2}\sqrt{\frac{\pi}{a}}
\mbox{erfi}\left( \sqrt{a} x \right).
\end{align}
\section{Derivation of the exact SER}\label{sec:app4}
The integral in \dref{eq:macro_mrc:Q2_expectation1} is required for
the exact SER analysis. Substituting $F_{\gamma} \left(w^2
/b\right)$ from \dref{eq:mrc:cdf2} into
\dref{eq:macro_mrc:Q2_expectation1} gives two new integrals
involving $F^1_{ik} \left(w^2 /b\right)$ or $F^2_{ik} \left(w^2
/b\right)$, which are given in \dref{eq:mrc:app3:cdf12} and
\dref{eq:mrc:app3:cdf22}. These two integrals can be written in
terms of known functions and two fundamental probability integrals
that we denote $I_{1} \left( \alpha , \beta \right)$ and $I_{2}
\left( \alpha , \beta \right)$. These integrals are computed below.

\subsection{Integral Form I} \label{subsec:4A}
Consider the integral,
\begin{align}
I_{1} \left( \alpha , \beta \right) &= \int_0^\infty x e^{-\beta x^2
}Q\left(x\right) \Phi \left( j\alpha x \right) dx.
\label{eq:mrc:integral_form1}
\end{align}
Applying the integral forms of $Q(.)$ and $\Phi(.)$ gives
\begin{align}
I_1 \left( \alpha , \beta \right) &= \frac{2j}{\pi}\int_0^\infty
\int_{\frac{x}{\sqrt{2}}}^\infty \int_0^{\alpha x} x e^{-\beta x^2
-t_1^2 + t_2^2} dt_2 dt_1 dx.
\end{align}
Using the substitutions, $t_1=r \cos\theta$ and $t_2=r \sin \theta$,
the integral then becomes
\begin{align}
I_1 \left( \alpha , \beta \right) &= \frac{2j}{\pi} \int_0^\infty
\int_{0}^\phi \int_{r_1}^{r_2} x r e^{-\beta x^2 - r^2cos\left.
2\theta\right.} dr d\theta dx,
\end{align}
where $\tan \phi = \alpha \sqrt{2}$, $r_1=x/\sqrt{2} \cos \theta$
and $r_2=\alpha x / \sin \theta$. Using standard methods of
integration with some simplifications we obtain
\begin{align}
\begin{split}
I_1 \left( \alpha , \beta \right) &= \frac{j}{2 \pi \beta}
\int_0^{\alpha \sqrt{2}} \frac{dt}{t^2 - 1 - 2\beta} \\
& \qquad \qquad - \frac{j\alpha^2}{2 \pi \beta} \int_0^{\alpha
\sqrt{2}} \frac{dt}{\left(\alpha^2 -\beta \right) t^2 - \alpha^2}.
\end{split}
\label{eq:mrc:integral_form11}
\end{align}
Defining
\begin{align}
I_{11} \left( \alpha , \beta \right) &= \int_0^{\alpha \sqrt{2}} \frac{dt}{t^2 - 1 - 2\beta}, \label{eq:mrc:integral_form111} \\
I_{12} \left( \alpha , \beta \right) &= \int_0^{\alpha \sqrt{2}}
\frac{dt}{\left(\alpha^2 -\beta \right) t^2 - \alpha^2},
\label{eq:mrc:integral_form112}
\end{align}
allows \dref{eq:mrc:integral_form11} to be rewritten as
\begin{align}
I_{1} \left( \alpha , \beta \right) &= \frac{j}{2 \pi \beta} I_{11}
\left( \alpha , \beta \right) - \frac{j\alpha^2}{2 \pi \beta} I_{12}
\left( \alpha , \beta \right).
\end{align}
The integral in \dref{eq:mrc:integral_form111} and
\dref{eq:mrc:integral_form112} can be solved in closed form to give
\begin{align}
I_{11} \left( \alpha , \beta \right) &= -\frac{1}{\sqrt{2\beta + 1}}
\tanh^{-1} \left( \frac{\alpha \sqrt{2}}{\sqrt{2\beta + 1}} \right)
\end{align}
and
\begin{align}
\!\!\!I_{12} \left( \alpha , \beta \right) \!\!&=\!\! \begin{cases}
-\frac{\sqrt{2}}{\alpha} & \quad \beta = \alpha^2 \\
-\frac{1}{\alpha \sqrt{\beta - \alpha^2}} \tan^{-1} \!\! \left(
\sqrt{2\left( \beta - \alpha^2 \right) } \right) & \quad
\mbox{otherwise}.
\end{cases}
\end{align}
Note that some intermediate steps in the derivation show that $1 +
2\beta
> 2\alpha^2$ is required for the existence of
\dref{eq:mrc:integral_form1}. This constraint is satisfied by the
current problem. This can easily be seen by substituting the
arguments of both $I_1\left(.,.\right)$ functions in
\dref{eq:mrc:Q2_ser2} in to $1 + 2\beta > 2\alpha^2$ followed by
simplifications using \dref{eq:mrc:app3:omega_alpha1}.
\subsection{Integral Form II} \label{subsec:4B}
Consider the integral,
\begin{equation}
I_{2} \left( \alpha , \beta \right) = \int_0^\infty x e^{\beta x^2
}Q\left(x\right) \left( 1 - \Phi \left( \alpha x \right) \right) dx.
\label{eq:mrc:integral_form2}
\end{equation}
Applying the integral forms of $Q(.)$ and $\Phi(.)$ we obtain
\begin{align}
I_2 \left( \alpha , \beta \right) &= \frac{2}{\pi}\int_0^\infty
\int_{\frac{x}{\sqrt{2}}}^\infty \int_{\alpha x}^{\infty} x e^{\beta
x^2 -t_1^2 - t_2^2} dt_2 dt_1 dx.
\end{align}

\noindent Following the same procedure as in Appendix
\ref{subsec:4A} and with some simplifications we arrive at
\begin{align}
\begin{split}
I_{2} \left( \alpha , \beta \right) = \frac{\alpha^2}{2 \pi \beta}
&\int_0^\phi \frac{d\theta}{ \alpha^2 - \beta \sin^2\theta } \\
&+ \frac{1}{2 \pi \beta} \int_\phi^{\pi/2} \frac{d\theta}{ 1 - 2
\beta \cos^2\theta } - \frac{1}{4\beta},
\label{eq:mrc:integral_form21}
\end{split}
\end{align}
where $\tan \phi=\sqrt{2}\alpha$. Making another substitution as
$t=\tan \theta$ in \dref{eq:mrc:integral_form21} gives
\begin{align}
\begin{split}
I_{2} \left( \alpha , \beta \right) = \frac{\alpha^2}{2 \pi \beta}
&\int_0^{\sqrt{2}\alpha} \frac{dt}{ \left(\alpha^2 - \beta
\right)t^2 + \alpha^2 } \\
&+ \frac{1}{2 \pi \beta} \int_{\sqrt{2}\alpha}^{\infty} \frac{dt}{
t^2 + 1 - 2 \beta } - \frac{1}{4\beta}.
\label{eq:mrc:integral_form22}
\end{split}
\end{align}

Defining
\begin{align}
I_{21} \left( \alpha , \beta \right) &= \int_0^{\sqrt{2}\alpha} \frac{dt}{ \left(\alpha^2 - \beta \right)t^2 + \alpha^2 }, \label{eq:mrc:integral_form211} \\
I_{22} \left( \alpha , \beta \right) &=
\int_{\sqrt{2}\alpha}^{\infty} \frac{dt}{ t^2 + 1 - 2 \beta },
\label{eq:mrc:integral_form212}
\end{align}
allows \dref{eq:mrc:integral_form22} to be rewritten as
\begin{align}
I_{2} \left( \alpha , \beta \right) &= \frac{\alpha^2}{2 \pi \beta}
I_{21} \left( \alpha , \beta \right) + \frac{1}{2 \pi \beta} I_{22}
\left( \alpha , \beta \right) - \frac{1}{4 \beta}.
\end{align}
The integrals in \dref{eq:mrc:integral_form211} and
\dref{eq:mrc:integral_form212} can be solved in closed form to give
\begin{align}
\!\!\!I_{21} \left( \alpha , \beta \right) \!\!&=\!\! \begin{cases}
\frac{\sqrt{2}}{\alpha} &  \quad \alpha^2 \!=\! \beta \\
\frac{1}{\alpha \sqrt{\alpha^2-\beta}} \tan^{-1} \!\left(\!
\sqrt{2\left( \alpha^2 \!-\! \beta \right) } \right) & \quad
\mbox{otherwise}
\end{cases}
\label{eq:mrc:integral_form211_solution}
\end{align}
and
\begin{align}
\!\!\! I_{22} \left( \alpha , \beta \right) &= \begin{cases}
\frac{1}{\sqrt{2} \alpha} &  \quad 1 = 2\beta \\
\frac{1}{ \sqrt{2\beta-1}}\coth^{-1} \left(
\frac{\sqrt{2}\alpha}{\sqrt{\left. 2\beta -1 \right.}} \right) &
\quad \mbox{otherwise}.
\end{cases}
\label{eq:mrc:integral_form212_solution}
\end{align}
As for $I_{1} \left( \alpha , \beta \right)$ there is an associated
region of validity, $1 + 2\alpha^2 > 2 \beta$, which is satisfied by
the problem.
%
%

%\begin{IEEEbiography}[{\includegraphics[width=1in,height=1.25in,clip,keepaspectratio]{./TCOM-TPS-12-0242_R2_dushyantha.jpg}}]{Dushyantha Basnayaka}
\begin{IEEEbiography}{Dushyantha Basnayaka}
(S'11-M'12) was born in 1982 in Colombo, Sri Lanka. He received the
B.Sc.Eng degree with 1\textsuperscript{st} class honors from the
University of Peradeniya, Sri Lanka, in Jan 2006. He is currently
working towards for his PhD degree in Electrical and Computer
Engineering at the University of Canterbury, Christchurch, New
Zealand.\\
He was an instructor in the Department of Electrical and Electronics
Engineering at the University of Peradeniya from Jan 2006 to May
2006. He was a system engineer at MillenniumIT (a member company of
London Stock Exchange group) from May 2006 to Jun. 2009. Since Jul.
2009 he is with the communication research group at the University
of Canterbury, New Zealand.\\
D. A. Basnayaka is a recipient of University of Canterbury
International Doctoral Scholarship for his doctoral studies at UC.
His current research interest includes all the areas of digital
communication, especially macrodiversity wireless systems. He holds
one pending US patent as a result of his doctoral studies at UC.
\end{IEEEbiography}
%
%
%\begin{IEEEbiography}[{\includegraphics[width=1in,height=1.25in,clip,keepaspectratio]{./TCOM-TPS-12-0242_R2_pete.jpg}}]{Peter Smith}
\begin{IEEEbiography}{Peter Smith}
(M'93-SM'01) received the B.Sc degree in Mathematics and the Ph.D
degree in Statistics from the University of London, London, U.K., in
1983 and 1988, respectively. From 1983 to 1986 he was with the
Telecommunications Laboratories at GEC Hirst Research Centre. From
1988 to 2001 he was a lecturer in statistics at Victoria University,
Wellington, New Zealand. Since 2001 he has been a Senior Lecturer
and Associate Professor in Electrical and Computer Engineering at
the University of Canterbury in New Zealand. Currently, he is a full
Professor at the same department.\\
His research interests include the statistical aspects of design,
modeling and analysis for communication systems, especially antenna
arrays, MIMO, cognitive radio and relays.
\end{IEEEbiography}
%
%
%\begin{IEEEbiography}[{\includegraphics[width=1in,height=1.25in,clip,keepaspectratio]{./TCOM-TPS-12-0242_R2_phil.jpg}}]{Philippa Martin}
\begin{IEEEbiography}{Philippa Martin}
(S'95-M'01-SM'06) received the B.E. (Hons. 1) and Ph.D. degrees in
electrical and electronic engineering from the University of
Canterbury, Christchurch, New Zealand, in 1997 and 2001,
respectively.  From 2001 to 2004, she was a postdoctoral fellow,
funded in part by the New Zealand Foundation for Research, Science
and Technology (FRST), in the Department of Electrical and Computer
Engineering at the University of Canterbury.  In 2002, she spent 5
months as a visiting researcher in the Department of Electrical
Engineering at the University of Hawaii at Manoa, Honolulu, Hawaii,
USA.  Since 2004 she has been working at the University of
Canterbury as a lecturer and then as a senior lecturer. Currently,
she is an Associate Professor at the same department. In 2007, she
was awarded the University of Canterbury, College of Engineering
young researcher award.  She served as an Editor for the IEEE
Transactions on Wireless Communications 2005-2008 and regularly
serves on technical program committees for IEEE conferences. \\
Her current research interests include multilevel coding, error
correction coding, iterative decoding and equalization, space-time
coding and detection, cognitive radio and cooperative communications
in particular for wireless communications
\end{IEEEbiography}

\end{document}